\documentclass[prr,twocolumn,amsmath,amssymb,superscriptaddress]{revtex4-1}
\usepackage{graphicx}
\usepackage{amsthm}
\usepackage{array}
\usepackage{hyperref}
\usepackage{xcolor}
\usepackage{mathtools}
\usepackage{bm,times,enumitem}
\hypersetup{colorlinks=true, urlcolor={black}, linkcolor={blue!80!black}, citecolor={blue!80!black}}

\newcommand{\ket}[1]{\left|{#1}\right>}
\newcommand{\bra}[1]{\left<{#1}\right|}
\newcommand{\tr}[1]{\mathrm{Tr}\!\left[#1\right]}
\newcommand{\sket}[1]{|{#1}\rangle\!\rangle}
\newcommand{\sbra}[1]{\langle\!\langle{#1}|}
\newcommand{\sinner}[2]{\langle\!\langle{#1}|{#2}\rangle\!\rangle}

\begin{document}
\title{Universal compressive characterization of quantum dynamics}

\author{Yosep Kim}
\thanks{These two authors contributed equally.}
\affiliation{Department of Physics, Pohang University of Science and Technology (POSTECH), 37673 Pohang, Korea}
\author{Yong Siah Teo}
\email{ys\_teo@snu.ac.kr}
\affiliation{Department of Physics and Astronomy, Seoul National University, 08826 Seoul, Korea}
\author{\!\!\normalfont\textsuperscript{, $\ast$} Daekun Ahn}
\affiliation{Department of Physics and Astronomy, Seoul National University, 08826 Seoul, Korea}
\author{Dong-Gil Im}
\affiliation{Department of Physics, Pohang University of Science and Technology (POSTECH), 37673 Pohang, Korea}
\author{Young-Wook Cho}
\affiliation{Center for Quantum Information, Korea Institute of Science and Technology (KIST), 02792 Seoul, Korea}
\author{Gerd~Leuchs}
\affiliation{Max-Planck-Institut f\"ur die Physik des Lichts, Staudtstra\ss e 2, 91058 Erlangen, Germany}
\affiliation{Institute of Applied Physics, Russian Academy of Sciences, 603950 Nizhny Novgorod, Russia}
\author{Luis~L.~S{\'a}nchez-Soto}
\affiliation{Max-Planck-Institut f\"ur die Physik des Lichts, Staudtstra\ss e 2, 91058 Erlangen, Germany}
\affiliation{Departamento de \'Optica, Facultad de F\'{\i}sica, Universidad Complutense, 28040 Madrid, Spain}
\author{Hyunseok Jeong}
\email{jeongh@snu.ac.kr}
\affiliation{Department of Physics and Astronomy, Seoul National University, 08826 Seoul, Korea}
\author{Yoon-Ho Kim}
\email{yoonho72@gmail.com}
\affiliation{Department of Physics, Pohang University of Science and Technology (POSTECH), 37673 Pohang, Korea}

\date{\today}

\begin{abstract}
	Recent quantum technologies utilize complex multidimensional processes that govern the dynamics of quantum systems. We develop an adaptive diagonal-element-probing compression technique that feasibly characterizes any unknown quantum processes using much fewer measurements compared to conventional methods. This technique utilizes compressive projective measurements that are generalizable to arbitrary number of subsystems. Both numerical analysis and experimental results with unitary gates demonstrate low measurement costs, of order $O(d^2)$ for $d$-dimensional systems, and robustness against statistical noise. Our work potentially paves the way for a reliable and highly compressive characterization of general quantum devices.
\end{abstract}
\maketitle

{\it Introduction.---}Quantum processes are nature's directives that guide the evolution of all physical systems in the quantum realm. Such processes ubiquitously occur in untamed open-system dynamics under interactions with the environment as, for example, depolarizing~\cite{Jeong:2013aa}, dephasing~\cite{Lim:2015aa} and photon-loss~\cite{Kim:2012aa} channels. They also exist as universal gates to carry out quantum computation. Notably, quantum processors~\cite{Ladd:2010aa,Campbell:2017aa,Ladd:2010aa,Lekitsch:2017aa} employ a series of such unitary processes~\cite{Schafer:2018aa,Shi:2018aa,Ono:2017aa,Patel:2016aa,Fiurasek:2008fg} to carry out computations using $d$-dimensional systems as resources. Thus, reliable characterizations of quantum processes are crucial prerequisites for enhancing the quality of quantum technologies. Such a characterization conventionally require $O(d^4)$ measurements~\cite{Chuang:2000fk,OBrien:2004aa,Fiurasek:2001dn,Poyatos:1997aa,Teo2011aa} that are too resource-intensive to perform for large $d$. Ancilla-~\cite{Altepeter:2003aa,Leung:2003aa,D'Ariano:2003aa,D'Ariano:2001aa} and error-correction-based~\cite{Omkar:2015aa,Omkar:2015qc,Mohseni:2007aa,Mohseni:2006aa} quantum process tomography (QPT) were introduced to circumvent this problem. These demand sophisticated state and measurement preparations. For specific property prediction tasks, direct schemes may be sufficient~\cite{Kim:2018sw,Gaikwad:2018aa,Bendersky:2013aa,Schmiegelow:2011aa,Bendersky:2009aa,Bendersky:2008aa}.

When the unknown process has a certain maximum possible rank, the concept of compressed sensing~\cite{Donoho:2006cs,Candes:2006cs,Candes:2009cs,Gross:2010cs,Kalev:2015aa,Steffens:2017cs,Riofrio:2017cs} has so far been the \emph{status quo} for reconstructing the unknown process with a small set of specialized measurements~\cite{Baldwin:2014aa,Rodionov:2014aa,Shabani:2011aa}. In practice however, this concept is only as reliable as the accuracy of the rank knowledge, and lacks an independent verification method to check the reconstruction results without fidelity comparison with target processes~\cite{Rodionov:2014aa,Shabani:2011aa}. Existing remedies for tackling these issues in compressed sensing are generally \emph{ad hoc} and incomplete~\cite{Shchukina:2017cs}.

In what follows, we shall present and experimentally demonstrate an adaptive compressive quantum process tomography scheme (ACQPT) that uniquely characterizes any process through direct diagonal-element-probing measurements in optimally-chosen bases that are much fewer than $O(d^4)$, \emph{ergo} highly compressive. Our scheme does not rely on any sort of prior assumption about the process, with the exception that one knows the dimension $d$ of the underlying quantum system. Instead, it is designed to extract information that is already inherently encoded in the measured data to reveal all process-matrix elements and check if they are uniquely consistent with the data using an efficient semidefinite program~\cite{Boyd:2004qd,Vandenberghe:1996ca}. If not, the scheme adaptively chooses the next optimal measurement to perform, and repeats itself until the process is uniquely characterized. 

We shall elaborate the theoretical formalism of ACQPT, and demonstrate that it is both highly compressive and achievable in practice using a proof-of-principle quantum optics experiment for two-qubit processes. Numerical simulations of a range of dimensions supply compelling evidence of an $O(d^2)\ll O(d^4)$ measurement cost for characterizing qudit unitary processes, while experimental data confirms the robustness of ACQPT in the presence of statistical noise. %A separate and rather different scheme with optical fibers was also reported.\cite{Teo:2019of}

\begin{figure*}[t]
	\centering
	\includegraphics[width=1.95\columnwidth]{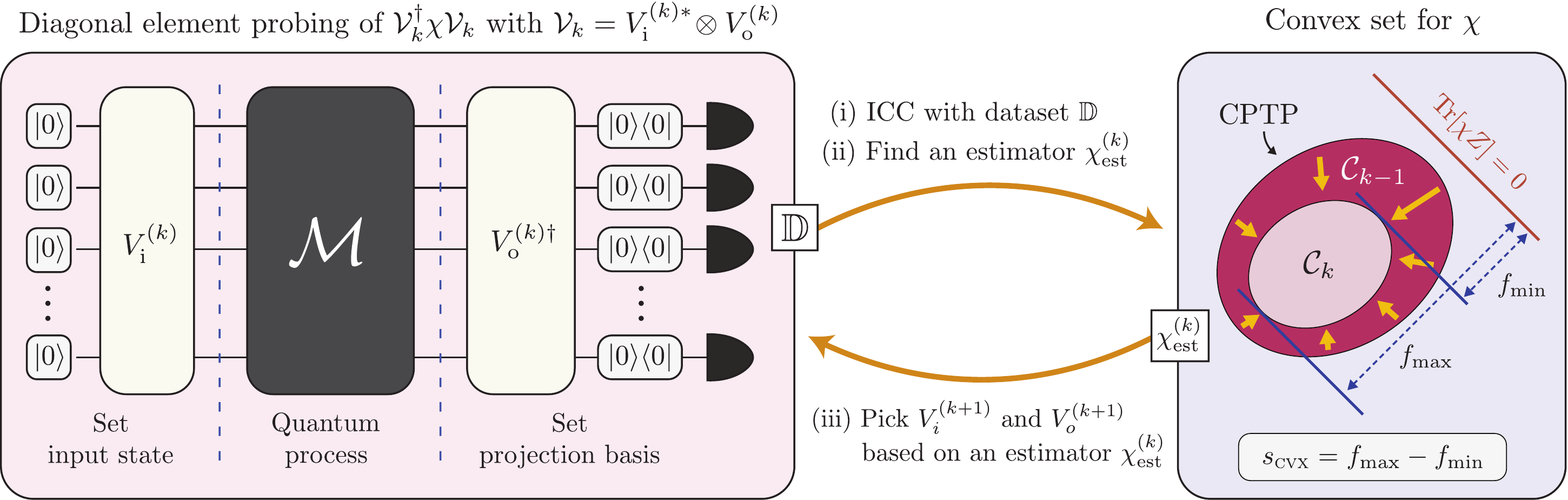}
	\caption{An iterative schematic of adaptive compressive quantum process tomography (ACQPT) to find out the $\chi$ matrix that represents a quantum process $\mathcal{M}$. This schematic is feasible in practice and generalizes to systems of any dimension, making ACQPT universal. At a particular step $k$, the input state and projection basis are set by the control unitaries $V^{(k)}_{\mathrm{i}}$ and $V^{(k)}_{\mathrm{o}}$ that defines the $\chi$-rotation of $\mathcal{V}_{k}^{\dagger}\chi\mathcal{V}_{k}$ with $\mathcal{V}_{k}=V^{(k)*}_\mathrm{i}\! \otimes V^{(k)}_\mathrm{o}$. The detection probability that corresponds to the first diagonal element is accumulated in the dataset $\mathbb{D}$. Because the diagonal element probing of $U_k^\dag\chi U_k$ in general demands resource-intensive measurements, $U_k^\dag\chi U_k$ is replaced by the closely approximated $\mathcal{V}_k^\dag\chi \mathcal{V}_k$ whose diagonal element can be measured with the simple experimental setup. Completely positive and trace-preserving process (CPTP) matrices $\chi$ that satisfy the dataset $\mathbb{D}$ form a convex set $\mathcal{C}_k$, which shrinks as data accumulate. In informational completeness certification (ICC), we track this shrinkage with a linear function $f(\chi)=\mathrm{Tr}[\chi Z]/\sqrt{\mathrm{Tr}[Z^2]}$, where $Z$ is a full-rank positive operator. Accordingly, the size monotone $s_\textsc{cvx}$ is defined as the difference between its unique maximum $f_\mathrm{max}$ and minimum $f_\mathrm{min}$ over $\mathcal{C}_k$. If $\mathcal{C}_k$ contains only a single estimator ($s_\textsc{cvx}=0$), then this unique estimator $\chi_{\mathrm{est}}^{(k)}$ is used to represent the quantum process and terminate ACQPT. If not, ACQPT picks an optimal estimator $\chi_{\mathrm{est}}^{(k)}$ having the minimum entropy from the convex set $\mathcal{C}_k$ to pick the next $V_i^{(k+1)}$ and $V_o^{(k+1)}$.}
	\label{fig:acqpt}
\end{figure*}

{\it Compressive characterization of physical processes.---}Every quantum process that governs natural phenomena can be completely described by a positive semidefinite matrix $\chi$, which is defined by $O(d^4)$ parameters~\cite{Chuang:2000fk}. This matrix represents a state-to-state transformation rule for the process that accounts for all physical characteristics of the quantum system. We shall unambiguously determine $\chi$ using minimal number of measurements necessary with no other presumed information apart from knowing the dimension $d$ of the system ($\mathrm{dim}\{\chi\}=d^2$). Without loss of generality, we shall investigate trace-preserving processes, examples of which include all unitary processes used in quantum computation. All subsequent discussions directly apply also to non-trace-preserving processes if so desired.

%\vspace{1ex}\noindent{\bf Theoretical formalism.}~
Characterizing a physical process is equivalent to unambiguously finding out the elements of $\chi$. The aim is to do this with as little measurement resources as possible without the need for any other \emph{a priori} information about $\chi$. We first emphasize a pivotal observation about rank-deficient $\chi$ matrices: When the unknown process is unitary, its $\chi$ matrix is rank-1 and possesses only one positive eigenvalue. Then, we just need to diagonalize $\chi$ \emph{via} its diagonalizing unitary matrix $U_\mathrm{diag}$ and measure that single eigenvalue to fully characterize $\chi$ (The trace-preserving condition would have fixed this eigenvalue anyway so no measurement is even needed in this case). This straightforward argument can be extended to a rank-$r$ process. In this case, we simply measure all $r-1$ positive diagonal values of $U_\mathrm{diag}^\dag\chi U_\mathrm{diag}$. In other words, diagonal-element measurements, in view of the prior knowledge about $U_\mathrm{diag}$, supply the most information compared to other kinds of measurements. 

Evidently, as one has no knowledge about $\chi$, $U_\mathrm{diag}$ is also unknown. Nonetheless, we can design ACQPT to adaptively choose a sequence of unitary rotations $\mathcal{U}=\{U_1,U_2,\ldots,U_k,\ldots\}$ on $\chi$ that converges to $U_\mathrm{diag}$. Put simply, the iterative scheme executes two basic stages per step. In stage~(i), the scheme deterministically certifies if the accumulated dataset $\mathbb{D}$ from the experiment correspond to a unique estimator $\chi_{\mathrm{est}}$ for the unknown $\chi$ matrix---the \emph{informationally complete} (IC) situation. The contrary would imply that there is a convex set of processes $\mathcal{C}$ consistent with the non-IC $\mathbb{D}$~\cite{Kosut:2004aa,Kosut:2009aa}. We call this the informational completeness certification (ICC) stage, and its successful implementation stems from the convexity property of $\mathcal{C}$ that allows us to assign a mathematically justified number $s_\textsc{cvx}$ (size monotone) to indicate whether $\mathbb{D}$ is IC ($s_\textsc{cvx}=0$) or not. 

For a realistic numerical approach, we preset a certain threshold $\varepsilon$ of $s_\textsc{cvx}$. If $s_\textsc{cvx}>\varepsilon$, ACQPT finds an optimal measurement setting to collect more data in stage~(ii) based on $\mathbb{D}$ alone. This is the adaptive measurement stage responsible for generating $\mathcal{U}$. Since our $\chi$ of interest is rank-deficient, we can turn ACQPT into a compressive scheme by employing an effective low-rank guiding prescription: After ICC, an estimator with the minimum von Neumann entropy (minENT) is chosen from $\mathcal{C}$~\cite{Ahn:2019aa,Ahn:2019ns,Huang:2016aa,Tran:2016aa}. The next optimal $U$ for the $\chi$-rotation would be the one that diagonalizes this estimator to be the eigenvalues in descending order. 

\begin{figure}[t]
	\centering
	\includegraphics[width=0.85\columnwidth]{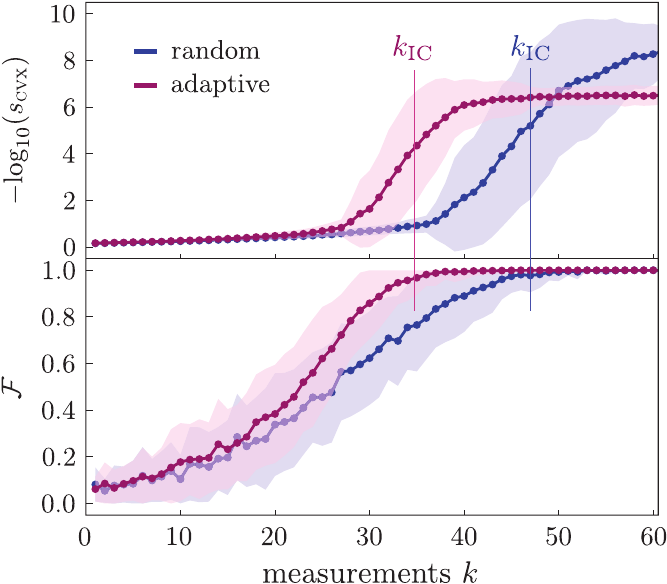}
	\caption{\label{fig:adapt_rand}Numerical simulations of adaptive and random strategies for $s_{\textsc{cvx}}$ and fidelity $\mathcal{F}$ with ACQPT for minimal nontrivial ququart unitary processes. A random rotation sequence is defined by Haar-random unitary matrices (see Appendix~\ref{app:supp_2}) and $\mathcal{F}$ in the $k$th iterative step is computed using the optimal $\chi_{\mathrm{est}}^{(k)}$ chosen with minENT. The markers and shaded regions indicate the average values and standard deviations respectively over 60 random ququart unitary processes. Here, the $k_\textsc{ic}$ values are recorded at the instant $s_{\textsc{cvx}}<5\times10^{-5}$, which are respectively $34.7\pm3.6$ and $47.0\pm5.9$ for the adaptive and random measurements. All simulations are statistically noiseless, so that $\mathcal{F}$ at $k_\textsc{ic}$ for \emph{every process} is unity to numerical precision.}
\end{figure}

At the $k$th iterative step, the $\kappa_{k}$th diagonal element of $U_{k}^\dag\chi U_{k}$ is measured following the rule $\kappa_{k}=\mathrm{mod}(k,r_{k-1})+1$, where $r_{k-1}=\mathrm{rank}\{\chi_{\mathrm{est}}^{(k-1)}\}$. The logic of this ``modulo rule'' is to measure the diagonal element of cyclically-shifted index within the positive-eigenvalue sector of the previously estimated $\chi_\mathrm{est}$, such that eventually all positive eigenvalues of $\chi$ are measured (up to statistical noise) at step $k=k_\textsc{ic}$, that is the final step at which IC measurement data are obtained. As an example, ACQPT measures the first diagonal element of $\chi$ at $k=1$, then the second diagonal element of $U_2^\dagger\chi U_2$ at $k=2$, and back to the first diagonal element of $U_3^\dagger\chi U_3$ if $r_2=\mathrm{rank}\{\chi_\mathrm{est}^{(2)}\}<3$, and so forth. 

Figure~\ref{fig:acqpt} shows an iterative schematic of ACQPT. The diagonal element probing of $U_k^\dag\chi U_k$ in general demands resource-intensive measurements, thus we approximate the probing in an experimental feasible way using a variable input $V^{(k)}_\mathrm{i}|0...0\rangle$ and projection onto $V^{(k)}_\mathrm{o}|0...0\rangle$. The first diagonal element of $\mathcal{V}_k^{\dagger}\chi\mathcal{V}_k$, where $\mathcal{V}_k=V^{(k)*}_\mathrm{i}\! \otimes V^{(k)}_\mathrm{o}$, can be obtained from the detection probability, and $V^{(k)}_\mathrm{i}$ and  $V^{(k)}_\mathrm{o}$ are chosen to closely approximate the $\kappa_{k}$th diagonal of $U_k^\dag\chi U_k$. We invite the reader to visit Appendix~\ref{app:supp_1} for further elaboration.

As ACQPT proceeds, more independent data are collected such that $U_k\rightarrow U_\mathrm{diag}$ quickly. In this way, our scheme can efficiently acquire optimal data and determine whether they are sufficient to uniquely recover $\chi$ without ever requiring spurious pre-experimental assumptions about $\chi$. Notice that the rank-deficiency of $\chi$ is not assumed here. A (nearly) full-rank $\chi$ unbeknownst to us would automatically result in a much slower convergence of ACQPT.

\begin{figure}[t]
	\centering
	\includegraphics[width=0.95\columnwidth]{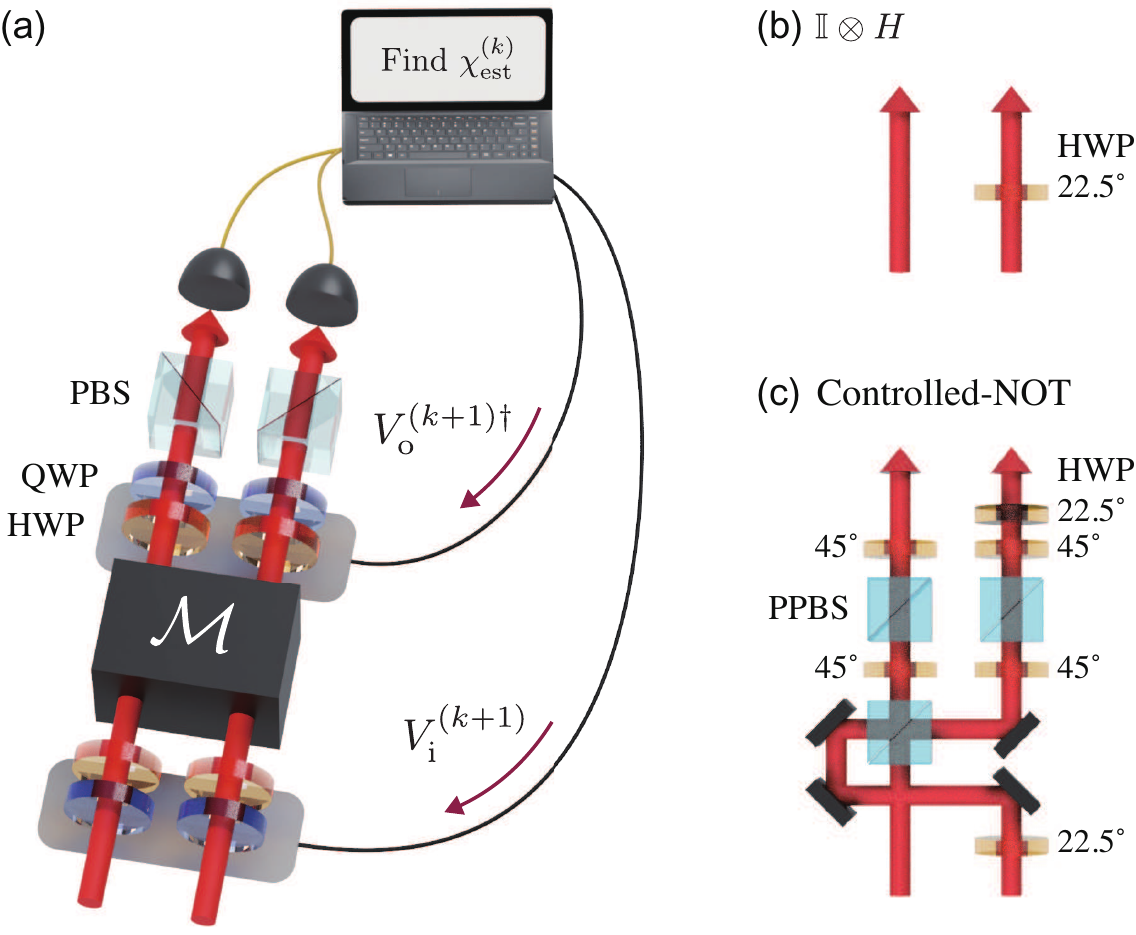}
	\caption{\label{fig:expt}(a) The polarization state of two-photons is transformed by a two-qubit process $\mathcal{M}$. For ACQPT on the unknown $\chi$ matrix that represents $\mathcal{M}$, the initial state and output projection basis are adaptively manipulated by controlling half-wave plates (HWP) and quarter-wave plates (QWP) according to $V_\mathrm{i}^{(k+1)}$ and $V_\mathrm{o}^{(k+1)}$, which are given from the previously estimated $\chi_{\mathrm{est}}^{(k)}$ based on the accumulated dataset (see Appendix~\ref{app:supp_1}). As the target two-qubit processes, $\mathbb{I}\otimes H$ (Hadamard transform only on the second qubit) and controlled-NOT (CNOT) gates are optically constructed. (b) A HWP at 22.5$^\circ$ is placed on the second qubit, which functions as the Hadamard gate. (c) Only if two single-photons have vertical polarization, Hong-Ou-Mandel interference on a partially polarizing beam splitter (PPBS) imprints a phase shift of $\pi$ onto two-qubit state, that is, controlled-phase gate. To make a CNOT gate, HWPs at 22.5$^\circ$ are applied on the second (target) qubit before and after the controlled-phase gate~\cite{Kiesel:2005aa,Okamoto:2005aa}.}
\end{figure}

\begin{figure*}[t]
	\centering
	\includegraphics[width=1.95\columnwidth]{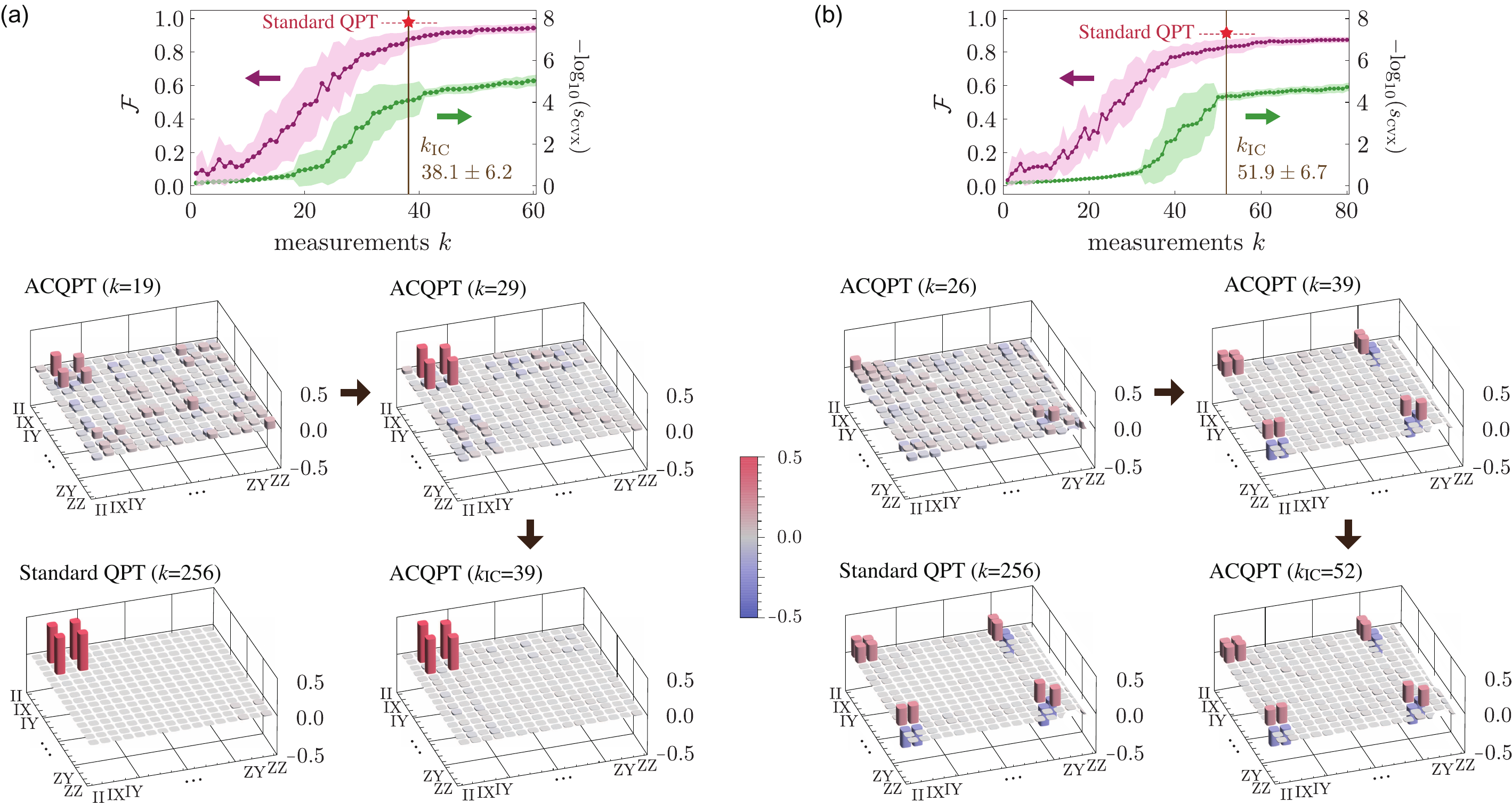}
	\caption{\label{fig:res}The experimental ACQPT results plotted for both the (a)~$\mathbb{I}\otimes H$ and (b)~CNOT gates. All $k_\textsc{ic}$ values are recorded at the instants when $s_{\textsc{cvx}}$ drops below $5\times10^{-5}$, and 15 experimental runs are performed on each gate. Although both gates should ideally be unitary, the CNOT gate possesses a $\chi$ matrix of higher rank due to the hardware imperfections such as polarization-dependent losses and partial distinguishability within a photon pair. Accordingly, the CNOT gate requires more measurements than the $\mathbb{I}\otimes H$ gate, and hence a larger $k_\textsc{ic}$ and a lower IC $\mathcal{F}$. The density-matrix plots of the minENT estimator $\chi_{\mathrm{est}}^{(k)}$ at various ACQPT step numbers $k$ for each gate paint the evolution picture of the compression run that leads to the final unique estimator $\chi_{\mathrm{est}}^{(k_\textsc{ic})}$, which is close to the standard QPT reconstruction using $k=256$ measurements.}
\end{figure*}

{\it Numerical results.---}Our first numerical showcase of ACQPT demonstrates the superiority of adaptive $\chi$-rotation sequences over random ones. For this, both the size monotone $s_{\textsc{cvx}}$ and fidelity $\mathcal{F}$ with the true process are numerically simulated with multiple randomly chosen ququart ($d=4$) unitary processes (see Fig.~\ref{fig:adapt_rand}). The results clearly show that the adaptive strategy gives a significantly smaller $k_\textsc{ic}$ than the random strategy. 

We further give numerical estimates on the scaling behaviors of $k_{\textsc{ic}}$ for both the adaptive and random strategies on qudit unitary processes in Appendix~\ref{app:supp_3}. We show, for a reasonably large range of $d$ and multiple simulations with random processes, that these strategies only need measurement resources of $O(d^2)$ in contrast with the standard $O(d^4)$. For a more complete analysis, we also compared the minENT strategy in ACQPT with another available adaptive strategy, the minimum-L1 norm strategy. 

{\it Experimental results.---}The experimental platform we use for demonstrating ACQPT utilizes a source of two-photons. The quantum processes of interest are implemented for the polarization modes. A 140-fs ultrafast laser is impinged on a 1 mm-thick type-II Barium Borate~(BBO) crystal to emit two single-photons in the beamlike configuration using spontaneous parametric down conversion (SPDC). The individual photons each possesses the central wavelength of 780 nm and delivered to the experimental setup shown in Fig.~\ref{fig:expt}(a) with single-mode fibers. To maximize the indistinguishability between two photons, their spectra are truncated by making use of interference filters having 2 nm full-width at half-maximum bandwidth.

The two-photon state is first initialized to the doubly horizontally-polarized state $\ket{00}\bra{00}$.  At the $k$th iterative step of ACQPT, the initial state and projection basis are manipulated using half- (HWP) and quarter-wave (QWP) plates according to the previously chosen two-qubit operations $V^{(k)}_\mathrm{i}$ and $V^{(k)}_\mathrm{o}$ respectively. For simple implementation, the closest separable initial state and projection basis are utilized. We anticipate an even better performance from ACQPT with sophisticated entangling operations.

The ACQPT proceeds as follows: Starting with $k=1$, $V^{(1)}_\mathrm{i}$ and $V^{(1)}_\mathrm{o}$ are generated randomly because there is initially no information about the unknown $\chi$. Making use of the algorithm in Appendix~\ref{app:supp_1}, a datum $\mathbb{D}_1$ is obtained from the probability of coincidentally detecting a photon at each of the two photodetectors [see Fig.~\ref{fig:expt}(a)]. This probability is estimated from dividing the coincidence counts for the setting of $V^{(1)}_\mathrm{i}$ and $V^{(1)}_\mathrm{o}$ by the total input photon counts, the latter which can be measured by removing the polarizing beam splitter (PBS) at once. The datum is then utilized to find $V^{(2)}_\mathrm{i}$ and $V^{(2)}_\mathrm{o}$ to measure in the next iteration, and the whole procedure repeats until a unique process matrix is completely characterized at $k=k_\textsc{ic}$. 

We investigate both the $\mathbb{I}\otimes H$ (Hadamard transform only on the second qubit) and controlled-NOT (CNOT) gates constructed as in Figs.~\ref{fig:expt}(b) and (c). The Hadamard gate is  simply realized using a HWP at 22.5$^\circ$, whereas the CNOT gate is implemented by exploiting Hong-Ou-Mandel interference effects on a partially polarizing beam splitter (PPBS) that partially reflects vertical polarization with a transmittance of 1/3 and perfectly transmits horizontal polarization~\cite{Kiesel:2005aa,Okamoto:2005aa,Hong:1987aa}. Figure~\ref{fig:res} shows plots of $s_{\textsc{cvx}}$ and $\mathcal{F}$ at each step $k$ for the two-qubit processes. ACQPT essentially gives almost the same results as standard QPT with much fewer measurement outcomes (38.1$\pm$6.2 and 51.9$\pm$6.7 for $\mathbb{I}\otimes H$ and CNOT gates) than $256$ and no prior information. The disparity in the required number of measurements for the two gates stems from their different degrees of implementation imperfections that result in non-unitary processes. This clearly shows that ACQPT works adaptively.

{\it Discussion.---}All presented simulation and experimental results have confirmed, indeed, that our adaptive element-probing compressive scheme can characterize any quantum process using drastically less measurement resources than the standard $O(d^4)$ without imposing \emph{ad hoc} assumptions. Additional simulation graphs and procedures illustrated in Appendix~\ref{app:supp_3} provide evidence that for general qudit unitary processes, there exists a quadratic enhancement to $O(d^2)$ in terms of measurement resource costs needed to unambiguously characterize any qudit unitary process in contrast with $O(d^4)$.

One may additionally incorporate trusted prior information into ACQPT. Most straightforwardly, if one insists in knowing the rank $r$ of the actual unknown process, then one may simply replace $r_k$ with $r$ in the ``modulo rule'' when measuring diagonal elements. Numerically for example, if we enforce $r=1$, ACQPT becomes comparable to the most efficient Baldwin-Kalev-Deutsch (BKD) scheme~\cite{Baldwin:2014aa,Finkelstein:2004aa} for unitary channel known to date, that requires projective measurements of $2d^2-d$. See Appendix~\ref{app:supp_3} for details. The advantage of incorporating the prior information this way into ACQPT is that even if the prior information turns out to be inaccurate, the effect is not detrimental since an additional layer of certification is carried out to verify if the process estimator is truly unique. This failsafe is what distinguishes ACQPT in merit from all other reported undersampled characterization schemes to the authors' knowledge. 

Furthermore, as shown in Fig.~\ref{fig:adapt_rand}, the size monotone and fidelity progress very similarly for both the adaptive and random strategies during the initial measurement phase. We may then use this observation to arrive at a hybrid compressive scheme where random measurements are first used at the initial phase before they are switched to adaptive ones. This would also further reduce the overall computational load in trying to execute the adaptive stage repeatedly.

\begin{acknowledgments}
	This work was supported in part by the National Research Foundation of Korea (NRF) (Grant Nos. 2019R1A2C3004812, 2019M3E4A1080074, 2019R1H1A3079890, 2020R1A2C1008609 and No. 2019R1A6A1A10073437), KIST institutional program (Project No. 2E30620), and the Spanish MINECO (Grant Nos. FIS201567963-P and PGC2018-099183-B-I00). Y.K. acknowledges support from the Global Ph.D. Fellowship by the NRF (Grant No. 2015H1A2A1033028).
\end{acknowledgments}

\appendix
\section{Theory and algorithm for adaptive compressive and assumption-free quantum process characterization} 
\label{app:supp_1}

\subsection{Statistically noiseless case}

A quantum process $\mathcal{M}$ that maps a $d$-dimensional quantum state $\rho$ to another state $\rho'=\mathcal{M}[\rho]$ according to
\begin{equation}
\mathcal{M}[\rho]=\sum_{m,n=1}^{d^2}\chi_{mn}B_{m}\rho B_{n}^{\dagger}\,,
\label{eq:qp}
\end{equation}
where $\{B_{m}\}$ is a set of Hermitian basis operators that are mutually trace-orthonormal ($\mathrm{Tr}\!\left[B_{m}^\dagger B_{n}\right]=\delta_{mn}$)~\cite{Chuang:2000fk}. In this operator basis, $\mathcal{M}$ has a matrix representation $\chi$ defined by the elements $\chi_{mn}$ in some computational basis. We hereby consider (with no loss of generality) completely positive and trace-preserving (CPTP) processes, where the positive process matrix $\chi\geq0$ satisfies an additional operator constraint
\begin{equation}
\sum_{m,n=1}^{d^2}\chi_{mn}B_n^{\dagger}B_m=\mathbb{I}
\label{eq:tp}
\end{equation}

To characterize any $d^2$-dimensional quantum physical process, one needs to uniquely determine all its $d^4-d^2$ independent parameters. The relevant dataset $\mathbb{D}$ collected in the experiment for this purpose approximately estimates the actual detection probabilities $p_k=\mathrm{Tr}\!\left[O_k\mathcal{M}[\rho_k]\right]$ that are accumulated from studying $\mathcal{M}$ with various pairs ($\rho_k,O_k$) of input states $\rho_k$ and output observables $O_k$ for the $k$th measurement chosen according to some prescription that shall be discussed shortly. This dataset, \emph{in the absence of statistical noise}, thus has a linear relationship with $\bm{\chi}$, a column of $\chi_{mn}$ elements, inasmuch as
\begin{equation}
\bm{p}=\Phi\bm{\chi}\,, 
\label{eq:lin_data}
\end{equation}
where the $K\times d^4$ transformation matrix possesses elements $\Phi_{k, mn}=\mathrm{Tr}\!\left[O_k B_{m}\rho_k B_{n}^{\dagger}\right]$. An informationally complete (IC) and noiseless $\mathbb{D}$ means that the linear data constraint in Eq.~\eqref{eq:lin_data} and CPTP operator constraints permit a unique characterization of the $\chi$ matrix. Such an IC situation can occur even when $K\equiv k_\textsc{ic}\ll d^4$.

The adaptive compressive quantum process tomography (ACQPT) scheme is designed to characterize $\chi$ (or $\bm{\chi}$) with much fewer than $O(d^4)$ of $p_{k}$s and requires no additional assumptions about the process. It progresses as an iterative procedure, where in each step it carries out informational completeness certification (ICC) to check if $\mathbb{D}$ is IC or not. If not, the scheme adaptively chooses the next optimal measurement to perform. 

As more data accumulate, the convex set $\mathcal{C}$ of all CPTP operators that obey Eq.~\eqref{eq:lin_data} eventually shrinks to a singleton that contains $\chi$ in the absence of statistical noise. To indicate if $\mathcal{C}$ is a singleton or not, we can define an indicator $s_{\textsc{cvx}}$ over $\mathcal{C}$ by first defining a linear function $f(\chi)=\mathrm{Tr}\!\left[\chi Z\right]/\sqrt{\mathrm{Tr}\!\left[Z^2\right]}$, which is a distance from $\chi$ to a hyperplane $\mathrm{Tr}\!\left[\chi Z\right]$ for any positive operator $Z$. Upon denoting the minimum and maximum of $f$ over $\mathcal{C}$ by $f_\mathrm{min}$ and $f_\mathrm{max}$ respectively, we may define the indicator $s_{\textsc{cvx}}= f_\mathrm{max}-f_\mathrm{min}$. It is well-known in the study of convex optimization that minimizing or maximizing such a linear $f$ over any convex region gives a unique optimum, so that the convexity of $\mathcal{C}$ immediately implies that $s_{\textsc{cvx}}=0\leftrightarrow\mathcal{C}=\{\chi\}$~(noiseless case). For a more realistic numerical approach, the instant $s_{\textsc{cvx}}$ reaches below a certain preset threshold, the present process matrix $\chi_{\mathrm{est}}^{(k)}$ is considered as the desired unknown target matrix and ACQPT is terminated. When the dataset $\mathbb{D}$ becomes IC, $s_{\textsc{cvx}}$ is abruptly reduced by a few orders. Thus, the threshold can be distinctly decided this way in practice.

The indicator $s_{\textsc{cvx},k}\geq0$ is in fact a \emph{size monotone} for the data convex set $\mathcal{C}_k$, which is a (non-strict) monotonically increasing function with the size $s_k$ of $\mathcal{C}_k$. In principle, this special property holds for any convex/concave function $f$ that defines this indicator. We can instructively prove this for a concave function $f(\chi)$. In this case, we have $f_{\text{max},1}\geq f_{\text{max},2}\geq\ldots$ and $f_{\text{min},1}\leq f_{\text{min},2}\leq\ldots$ It is clear that if $f_{\text{max},k+1}-f_{\text{min},k+1}<f_{\text{max},k}-f_{\text{min},k}$, then $\mathcal{C}_{k+1}\subset\mathcal{C}_{k}$. It follows immediately that if $s_{\textsc{cvx},k}\equiv(f_{\text{max},k}-f_{\text{min},k})/(f_{\text{max},1}-f_{\text{min},1})$, then $s_{\textsc{cvx},k}$ is a size monotone that decreases with increasing $k$. When $s_{\textsc{cvx},k_\textsc{ic}}=0$, the convexity of $\mathcal{C}_{k_\textsc{ic}}$ implies that $\mathcal{C}_{k_\textsc{ic}}$ must contain only $\chi$ due to the unique maximum possessed by $f$. Similar arguments hold for a convex $f$. Since $f(\chi)$ is a linear function, it can also be used to formulate the size monotone as it facilitates the class of semidefinite programs known to give unique stationary points in the CPTP space.

We emphasize that the above arguments hold provided that $Z$ is non-pathological. A trivial pathological instance would be when $Z=\mathbb{I}/d$, such that $f=1$ over $\mathcal{C}$. Another example is when $Z$ is rank-deficient, in which case any $\chi'\in\mathcal{C}$ (if exists) that lies in the kernel of $Z$ results in $f=0$ and $s_{\textsc{cvx}}=0\nleftrightarrow\mathcal{C}=\{\chi\}$ in general. A randomly-chosen full-rank $Z$ would therefore avoid such pathological situations.

After ICC, the adaptive measurement stage ensues if $s_\textsc{cvx}$ is not sufficiently small. We reiterate the basic observation that if an observer knows $U_\mathrm{diag}$ that diagonalizes a rank-$r$ $\chi$, then the rank-$r$ diagonalized $D=U_\mathrm{diag}^{\dagger}\chi U_\mathrm{diag}$ has $r$ nonzero parameters so that measuring all the $r-1$ independent diagonal terms can completely characterize the quantum process. In other words, the measurements of the diagonal terms of $D$ provide more information about the quantum channel than any other kind of measurements. Based on the above observation, we design the ACQPT scheme to make an informed guess about the unknown diagonalizing $U_\mathrm{diag}$ from the available dataset $\mathbb{D}$ at hand. Suppose that ACQPT now operates at the $k$th iterative step. Then since the unknown $\chi$ matrix is rank-deficient, the next optimal rotation $U_{k+1}$ that approximates $U$ is taken to be the one that diagonalizes a low-rank estimator $\chi_\mathrm{est}^{(k)}$ from $\mathcal{C}_k$, where the unknown rank $r$ is also estimated to be $r_k$, the rank of $\chi_\mathrm{est}^{(k)}$. The estimator $\chi_\mathrm{est}^{(k)}$ is defined to be the one that minimizes the entropy (minENT) over $\mathcal{C}_k$ which has been found to perform very well in compressive tomography.

After this optimization, we sort the $r_k$ estimated diagonal elements of the diagonal matrix $D_\mathrm{est}^{(k)}=U^\dag_{k+1}\chi_\mathrm{est}^{(k)}U_{k+1}$ in descending order and ensure that $U_{k+1}$ precisely gives the same sorted order of eigenvalues, before rotating $\chi$ with it in the ($k+1$)th step. After which, we measure the actual diagonal elements of $D^{(k+1)}=U^\dag_{k+1}\chi U_{k+1}$ using this sorted list as a guide. The straightforward logical action now is to measure the $[\kappa_{k+1}=\mathrm{mod}(k,r_k)+1]$-th largest diagonal term of $D^{(k+1)}$, which spreads the measurements over the predicted support of $\chi$, so that its entire actual support is covered with larger probability. As more linearly independent data are collected, the principle of tomography dictates that $\mathcal{C}_k\rightarrow\{\chi\}$, $U_{k+1}\rightarrow U_\mathrm{diag}$ and $r_k\rightarrow r$ as $k\rightarrow k_\textsc{ic}$, and the minENT adaptation compressively reduces the value of $k_\textsc{ic}$. 

\subsection{Statistically noisy case}

There are only a few easy adjustments to accommodate statistical noise in actual experiments. We first understand that this time, $\mathbb{D}$ is now a set of normalized detection counts that are no longer the actual probabilities. Hence, the column $\bm{p}$ on the left-hand side of Eq.~\eqref{eq:lin_data} must be replaced by another column of physical probabilities $\bm{p}_{\mathrm{est}}$ such that there shall still exist a CPTP solution $\bm{\chi}_{\mathrm{est}}$ for
\begin{equation}
\bm{p}_{\mathrm{est}}=\Phi\bm{\chi}_{\mathrm{est}}\,.
\label{eq:lin_data2}
\end{equation}
Now, for our experiment, the photon source generates photon counts that well follow a Poisson distribution every time we collect a particular datum $\mathbb{D}_k$. For large number of sampling copies, the distribution at each $k$ can then be further approximated to a Gaussian distribution with mean and variance both proportional to $p_k$. In statistics, there exists a \emph{log-likelihood function}, which is the logarithmic conditional probability 
\begin{equation}
\log L(\bm{\nu}|\bm{p})=-\sum^K_{k=1}\,\dfrac{(\nu_k-p_k)^2}{2p_k}
\end{equation}
of obtaining the observed normalized photon counts $\nu_k$ given the true probabilities $p_k$. For every $\bm{\nu}$ up to the ($k=K$)th measurement, we may obtain the physical probability column $\bm{p}_{\mathrm{est}}=\bm{p}_{\textsc{ml}}$ that approximates the unknown true detection probability column $\bm{p}$ by maximizing $\log L(\bm{\nu}|\bm{p}')$ over $\bm{p}'$ subject to the CPTP constraints---the \emph{maximum-likelihood (ML) method}. In effect, we have searched for the most likely physical probability column $\bm{p}_\textsc{ml}$ that can give rise to the observed normalized photon-count column $\bm{\nu}$.

The second easy adjustment is to now regard $\mathcal{C}_k$ as the convex set of process matrices that give the same maximum value of $\log L$ for the accumulated dataset up to the $k$th step. These are the process matrices that lie in the domain of the plateau of $L$. When $k=k_\textsc{ic}$, the final unique ML estimator $\chi_\textsc{ml}$ shall also clearly be different from $\chi$. The distance between the two matrices can be further reduced either with more sampling copies or additional new measurements. Apart from these, we stress that the double implication $s_{\textsc{cvx},k_\textsc{ic}}=0\leftrightarrow\mathcal{C}_{k_\textsc{ic}}=\{\chi_\textsc{ml}\}$ is perfectly robust against noise in the sense that even if $\mathbb{D}$ is statistically noisy, $\mathcal{C}$ is \emph{always} convex and all arguments leading to the above double implication relies solely on this convexity property.

\subsection{Efficient and realistic augmentations}

When $\chi$ is expressible in the operator basis whose elements are of the form 
\begin{equation}
B_{m} =|i\rangle\langle j|\,, 
\label{eq:basis}
\end{equation}
where $\mathrm{Tr}\!\left[B_{m}^\dagger B_{n}\right]=\delta_{mn}$, the diagonal element $\chi_{mm}$ can be directly estimated from the detection probability $p_m=\mathrm{Tr}\!\left[O_i\mathcal{M}[\rho_j]\right]$ of the input state $\rho_j = |j\rangle\langle j|$ and the output projection observable $O_i =|i\rangle\langle i|$. However, if $\chi$ cannot be expressed with operator basis elements of such a form, then measurements of diagonal elements become complicated.

For simplicity, we first assume to know the identity of the diagonalizing unitary $U_\mathrm{diag}$ of $\chi$. To reveal the operator bases of $D=U_\mathrm{diag}^{\dagger}\chi U_\mathrm{diag}$, we may rewrite Eq.~(\ref{eq:qp}) in terms of a new operator basis, 
\begin{equation}
\mathcal{M}[\rho]=\sum_{n=1}^{d^2}D_{nn}B'_{n}\rho B'^\dagger_{n}\,,\quad B'_n=\sum_{m=1}^{d^2} U_{mn} B_{m}\,.\label{eq:qpdiag}
\end{equation}
It is evident that if the reference basis elements $B_n$ takes the form in \eqref{eq:basis}, then the diagonalizing operator basis elements $B'_n=\lambda^{(n)}_1|\phi^{(n)}_1\rangle \langle \phi'^{(n)}_1|+\lambda^{(n)}_2|\phi^{(n)}_2\rangle \langle \phi'^{(n)}_2|+\ldots$, where $\lambda^{(n)}_1\geq \lambda^{(n)}_2\geq\ldots$, typically possess multiple components in its singular-value decomposition. Since these components are mutually noncommuting, a simultaneous measurement of all of them in order to determine $D_{nn}$ in one experiment incurs intrinsic quantum uncertainties, which turns out to be a physically impossible task. As a physically realistic alternative to $B'_n$, the closest operator in the form of Eq.~(\ref{eq:basis}) is defined as $S_n=|b_n\rangle\langle a_n|\equiv |\phi^{(n)}_1\rangle\langle \phi'^{(n)}_1|$, the rank-1 component corresponding to the largest singular value $\lambda^{(n)}_1$. Measuring such a rank-1 component corresponds to a measurement values that approximate $D_{nn}$. These largest-singular-value principle is applied to approximately measure diagonal elements of any rotated $\chi$ in the course of an ACQPT run. This is the first necessary augmentation to the idealized ACQPT procedure.

The $\chi$ element for $S_n=|b_n\rangle\langle a_n|$ can be measured with the input state $\rho=|a_n\rangle\langle a_n|$ and output observable $O=|b_n\rangle\langle b_n|$. The second augmentation finds the nearest product observables for both $\rho$ and $O$ in every iterative step when one is dealing with many-body systems. There are many ways to do this, one of which is to express $\rho=V_\mathrm{i}|0\rangle\langle0|V_\mathrm{i}^\dag$ and $O=V_\mathrm{o}|0\rangle\langle0|V_\mathrm{o}^\dag$ in terms of a reference state $|0\rangle\langle0|$, and next respectively look for product unitary operators $V^{(\mathrm{prod})}_\mathrm{i}$ and $V^{(\mathrm{prod})}_\mathrm{o}$ that minimizes the operator norms $\|V^{(\mathrm{prod})}_\mathrm{i}-V_\mathrm{i}\|$ and $\|V^{(\mathrm{prod})}_\mathrm{o}-V_\mathrm{o}\|$ over the tensor-product unitary space. 

\subsection{Explicit procedure of ACQPT.}

We hereby state the complete algorithm of ACQPT that is applicable to real experiments:

\noindent
\rule{\columnwidth}{1pt}\\
\noindent
{\bf ACQPT}\\
\noindent
Set $\varepsilon$ to a small numerical value (say $\approx10^{-5}$) and choose an operator basis $\{B_{d i+j+1}=|i\rangle\langle j|\}$, where $i,j\in\{0,1,...,d-1\}$ and $\langle i|j\rangle=\delta_{ij}$, to represent the physical process $\mathcal{M}$ with a $d^2\times d^2$ $\chi$ matrix that obeys the CPTP constraints.  Start with $k=1$ and pick a random unitary $U_1$ and set $\kappa_1=1$. Compute the transformed operator basis element $B'_{\kappa_1}=\sum_{m=1}^{d^2} U_{m\kappa_1} B_{m}$, extract its largest singular-value component denoted by $|b_{1}\rangle\langle a_{1}|$, and measure the normalized sampled counts $\nu_1$ of the pair (input state $\rho_1=|a_1\rangle\langle a_1|$, projector $O_1=|b_1\rangle\langle b_1|$). Let $|a_1\rangle=V^{(1)}_{\mathrm{i}}|0\rangle$ and $|b_1\rangle=V^{(1)}_{\mathrm{o}}|0\rangle$ with respect to some reference pure state ($\ket{0}\equiv\ket{00}$ for our two-qubit experiments). The detection probability corresponds to the first diagonal element of $\mathcal{V}_1^{\dagger}\chi\mathcal{V}_1$ with $\mathcal{V}_1= V^{(1)*}_{\mathrm{i}}\!\otimes V^{(1)}_{\mathrm{o}}$. In this way, the closest approximation for the $\kappa_1$th diagonal element of $U_1^{\dagger}\chi U_1$ is accomplished. If the quantum system is many-body, measure instead the normalized counts of the nearest separable counterparts. Fix a randomly-generated full-rank $d^2\times d^2$ positive square matrix $Z$ that is of unit trace and define $\bm{\nu}=\nu_1$.
\begin{enumerate}
	\item\label{step:ml} \textbf{ML:} Find the physical ML probabilities $\widehat{p}_{\textsc{ml}}$ given the data $\bm{\nu}$.
	\item\label{step:icc} \textbf{ICC stage:} Compute the unique minimum and maximum values of $f(\chi)$ over $\chi\in\mathcal{C}_k$, that is, subject to the CPTP constraints of $\chi$ and $\widehat{p}_{\textsc{ml}}=\Phi\bm{\chi}$, where $\Phi_{k,mn}=\tr{O_k B_{m}\rho_k B_{n}^{\dagger}}$. Obtain $s_\textsc{cvx}$. If $s_\textsc{cvx}<\varepsilon$, terminate ACQPT, otherwise proceed to the next step.
	\item\label{step:adapt} \textbf{Adaptive stage:} Find the minENT estimator $\chi_{\mathrm{est}}^{(k)}$ over $\chi\in\mathcal{C}_k$ that minimizes the process entropy $-\tr{\chi\log \chi}$.
	\item\label{step:supp_desc} Diagonalize $\chi_{\mathrm{est}}^{(k)}$, sort its eigenvalues in descending order and find its correct diagonalizing unitary $U_{k+1}$ such that $U_{k+1}^\dag\chi_{\mathrm{est}}^{(k)} U_{k+1}$ gives the correct ordered eigenvalue matrix. Find out the rank $r_k$ of $\chi_{\mathrm{est}}^{(k)}$ and calculate $\kappa_{k+1}=\mod(k,r_k)+1$.
	\item Increase $k$ by one.
	\item\label{step:measure} Compute $B'_{\kappa_k}=\sum_{m=1}^{d^2} U_{m{\kappa_k}} B_{m}$, extract its largest singular-value component, denoted by $|b_{k}\rangle\langle a_{k}|$, and measure the normalized sampled counts $\nu_k$ of the pair $(\rho_k=|a_{k}\rangle\langle a_k|,O_k=|b_k\rangle\langle b_k|)$. Then the unitary operators $V_\text{i}^{(k)}$ and $V_\text{o}^{(k)}$ used to experimentally implement the $(\rho_k,O_k)$ pair are defined as $|a_k\rangle=V^{(k)}_\mathrm{i}|0\rangle$ and $|b_k\rangle=V^{(k)}_\mathrm{o}|0\rangle$. 
	\item Repeat Steps~\ref{step:ml}--\ref{step:measure} until $\mathbb{D}$ becomes IC.
\end{enumerate}
\vspace{-2ex}
\noindent
\rule{\columnwidth}{1pt}

\section{Random unitary operations and rank-deficient processes}
\label{app:supp_2}

We first provide a short numerical procedure that generates the random unitary matrices needed for the study of random compressive strategies. These matrices are distributed uniformly according to the Haar measure of the unitary group. 

\noindent
\rule{\columnwidth}{1pt}\\
\noindent
\textbf{Constructing a random $\bm{d^2\times d^2}$ Haar unitary matrix}
\begin{enumerate}
	\item Generate a random $d\times d$ matrix $A$ with entries independently and identically distributed according to the standard Gaussian distribution.
	\item Compute $Q$ and $R$ from the QR decomposition $A=QR$.
	\item Define $R_\text{diag}=\mathrm{diag}\{R\}$.
	\item Define $L=R_\text{diag} \oslash |R_\text{diag}|$ ($\oslash$ refers to the Hadamard division).
	\item Define $U_\mathrm{Haar}=QL$.
\end{enumerate}
\vspace{-2ex}
\noindent
\rule{\columnwidth}{1pt}\\

We next supply another general recipe that generates a distribution of random rank-$r$ $\chi$ matrices that are used to investigate the behavior of $k_\textsc{ic}$ against $r$. To do this, we state that the ``rank'' of a process is equivalent to the number of linearly independent Kraus operators $K_l$ used to represent this process. To see this, recall the process evolution relation 
\begin{equation}
\mathcal{M}[\rho]=\sum_lK_l\rho K_l^\dag=\sum^{d^2}_{m,n=1}\chi_{mn}B_m\rho B_n^\dag\,.
\label{eq:KB}
\end{equation}
To proceed with a more convenient notation, we define the superket of an operator $A$, $\sket{A}$, as a basis-dependent transposition map that transforms $A$ into an element in the $d^2$-dimensional complex vector space through \emph{matrix-column stacking}. The superbra is then the adjoint of the superket, $\sbra{A^\dag}=\sket{A}^\dag$, so that we have $\mathrm{Tr}\!\left[A^\dag B\right]=\sinner{A^\dag}{B}$. With this new machinery, Eq.~\eqref{eq:KB} turns into
\begin{equation}
\sum_l\sket{K_l\rho K_l^\dag}=\sum^{d^2}_{m,n=1}\sket{B_m\rho B_n^\dag}\chi_{mn}\,.
\label{eq:KB2}
\end{equation}
Using the formula $\sket{AXB}=B^\textsc{t}\otimes A\sket{X}$, and the fact that either Eq.~\eqref{eq:KB} or \eqref{eq:KB2} must hold for any $\rho$, we arrive at
\begin{equation}
\sum_lK_l^\dag\otimes K_l=\sum^{d^2}_{m,n=1}\chi_{mn}B_n^\dag\otimes B_m\,,
\end{equation}
which leads to
\begin{align}
\chi_{mn}=&\,\sum_l\mathrm{Tr}\!\left[B_n K_l^\dag\right]\mathrm{Tr}\!\left[K_l B_m^\dag\right]\nonumber\\
=&\,\sbra{B_m^\dag}\left(\sum_l\sket{K_l}\sbra{K_l^\dag}\right)\sket{B_n}\,.
\end{align}
So, $\chi_{mn}$ are the (transposed) matrix elements of $\sum_l\sket{K_l}\sbra{K_l^\dag}$ in the basis $\{\sket{B_n}\}$, whose rank $r$ equals the degree of linear independence of the set of rank-1 superoperators $\{\sket{K_l}\sbra{K_l^\dag}\}$ and is independent of the basis choice.

We therefore essentially require a simple procedure to generate a random set of $r$ linearly-independent Kraus operators $K_l$. Furthermore, the property $\sum^r_{l=1}K_l^\dag K_l=\mathbb{I}$ is to be preserved:

\noindent
\rule{\columnwidth}{1pt}\\
\noindent
\textbf{Generate} $\bm{r}$ \textbf{random} $\bm{d\times d}$ \textbf{Kraus matrices}
\begin{enumerate}
	\item Generate $r$ random $d\times d$ matrices $A_l$ with entries independently and identically distributed according to the standard Gaussian distribution.
	\item Compute $S=\sum^r_{l=1}A_l^\dag A_l$.
	\item Define the $r$ Kraus matrices as $K_l=A_lS^{-1/2}$.
\end{enumerate}
\vspace{-2ex}
\noindent
\rule{\columnwidth}{1pt}

\section{Numerical studies of ACQPT}
\label{app:supp_3}

\begin{figure}[t]
	\centering
	\includegraphics[width=0.85\columnwidth]{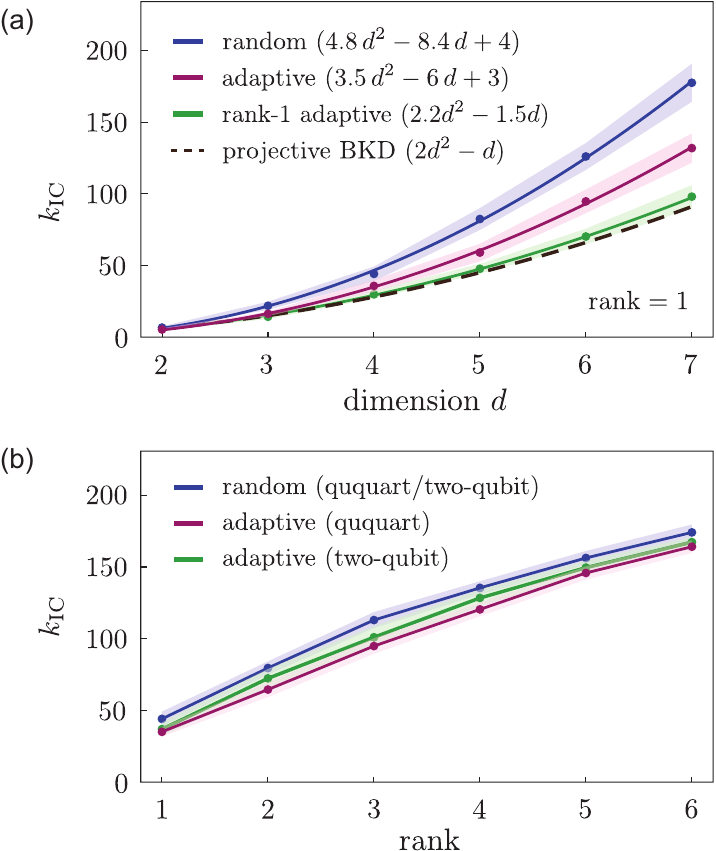}
	\caption{\label{fig:scaling}The respective asymptotic scaling behaviors of $k_{\textsc{ic}}$ for adaptive and random strategies against the (a) dimension $d$ and (b) rank are compared with the BKD unitary scheme using projective measurements, which almost matches the rank-1 adaptive strategy using the assumption $r_k=1$. The markers and shaded regions indicate the average values and standard deviations respectively over 30 random processes generated from the recipe in Sec.~II. The $k_\textsc{ic}$ values are recorded at the instant $s_{\textsc{cvx}}<5\times10^{-5}$.}
\end{figure}

In this section, we discuss the scaling behaviors of $k_{\textsc{ic}}$ for both ACQPT and the random strategy. For $d$-dimensional unitary processes, based on simulation results in the interval $2\leq d\leq 7$, we find that ACQPT can uniquely determine the process with an average of only about $3.5d^2$ measurements compared to the Haar-random strategy that requires $4.8d^2$ for large $d$, see Fig.~\ref{fig:scaling}. These results imply that both adaptive and random strategies that employ ICC for uniqueness certification use only $O(d^2)$ measurements, exponentially fewer than $O(d^4)$], to characterize qudit gates. For two-qubit processes, tensor-product local unitary rotation sequences are used on $\chi$ in ACQPT. Interestingly, ququart and two-qubit processes have almost identical $k_{\textsc{ic}}$ behaviors for high ranks.

\begin{figure}[t]
	\centering
	\includegraphics[width=0.85\columnwidth]{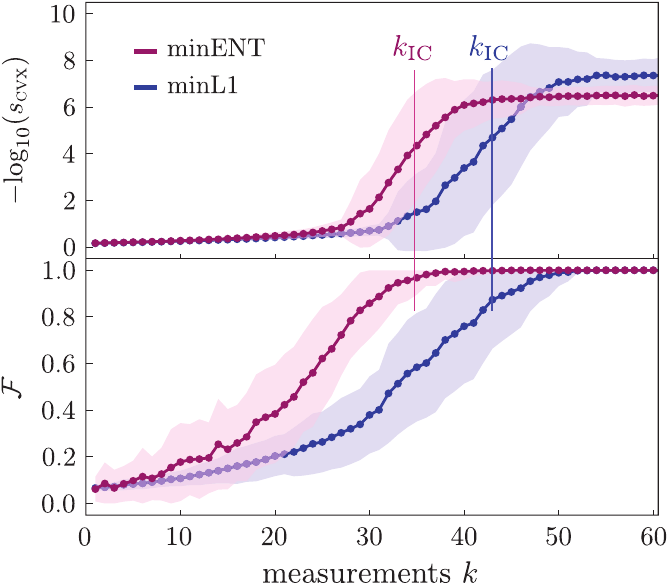}
	\caption{\label{fig:adapt_rand_minL1_vs_minent}Numerical simulations of ACQPT with two different optimization strategy: the default minENT and minL1 methods. The latter method picks $\chi_{\mathrm{est}}^{(k)}$ from $\mathcal{C}_k$ by minimizing the norm $\|U_k^\dag\chi U_k\|_1$ over all $\chi \in\mathcal{C}_k$ instead of entropy. The former is known to promote sparsity, which here refers to the (approximately found) diagonal basis representation of $\chi$. Graphs of $s_{\textsc{cvx}}$ and $\mathcal{F}$ are plotted against $k$ and averaged over 60 random ququart unitary processes. All other specifications are those for Fig.~2 in the main article. The $k_\textsc{ic}$ value for minL1 is $42.95 \pm 5.43$, which is larger than minENT, suggesting that ACQPT with minENT is still the more efficient compressive strategy.}
\end{figure}

We benchmark these two strategies with the known Baldwin-Kalev-Deutsch (BKD)~scheme for unitary channels that requires less \emph{projective} measurements~\cite{Baldwin:2014aa}. To derive the scaling behavior of the BKD scheme, Baldwin, Kalev and Deutsch argued that in order to characterize a process that is presumably unitary, one may choose to feed the process with a specific set of $d$ input pure states and characterize the corresponding output pure states. To reiterate their arguments, we parametrize the unknown unitary operator as $U=\sum^{d-1}_{j=0}|u_j\rangle\langle j|$, where $\langle u_j|u_k\rangle=\delta_{j,k}$ are kets we want to characterize and $\langle j|k\rangle=\delta_{j,k}$ are computational kets, and consider the set of $d$ input kets
\begin{align}
|\psi_0\rangle=&\,|0\rangle\,,\nonumber\\
|\psi_n\rangle=&\,(|0\rangle+|n\rangle)/\sqrt{2}\,,\quad 1\leq n\leq d-1\,.
\end{align}
Then feeding $|\psi_0\rangle$ to $U$ yields $|u_0\rangle$, and $|\psi_n\rangle$ gives $(|u_0\rangle+|u_n\rangle)/\sqrt{2}$. After $|u_0\rangle$ is fully determined, subsequent output states require the determination of fewer than $d$ amplitudes in the computational basis $\{|j\rangle\}$, the number of which decreases as $n$ increases. For instance, the output state corresponding to $\psi_{k}$ for some $n=k>0$ can be determined by characterizing the amplitudes of $|u_{k}\rangle$. Since all previous $k-1$ states are determined, we only need to characterize $d-k$ amplitudes $\langle j|u_k\rangle$ for $0\leq j\leq d-k-1$ and make use of $2k$ orthogonality relations with the kets $|u_0\rangle$ through $|u_{k-1}\rangle$. The total number of measurements needed is thus $k_\textsc{ic}=\sum^{d-1}_{k=0}(M-2k)=(M+1)d-d^2$, where $M$ is the number of outcomes needed to fully characterize an arbitrary pure state. 

In the original BKD unitary scheme, $M=2d$ is the number of outcomes of a complicated measurement scheme involving non-projective observables. In our context, we shall consider only rank-1 projective measurements that are more feasible to carry out in experiments. For this, there exists a lower bound of $M$ for such measurements, which is $3d-2$~\cite{Finkelstein:2004aa}. The final scaling behavior for the projective BKD unitary scheme then reads $k_\textsc{ic}=2d^2-d$.

We reiterate that these BKD optimal measurements, however, are effective only when the unknown process is strictly unitary. By contrast, ACQPT works without such a risky unitarity assertion and the reduction is also dramatic [$O(d^2)$] compared to $O(d^4)$ measurements employed in traditional QPT. Furthermore, the scaling behavior of the projective BKD scheme is, incidentally, very close to the IC number when $r_k$ is replaced by $r=1$ in the ``modulo rule'' used in ACQPT if one assumes that the unknown qudit process is unitary, which is $k_\textsc{ic}=O(2.2d^2)$ in the large-$d$ limit. This tells us that any presumed rank assumption can, and should be conservatively incorporated, in such a way that we allow ACQPT to decide if the data ultimately agree with such an assumption.

Figure~\ref{fig:adapt_rand_minL1_vs_minent} demonstrates the superiority of minENT over the minimum-L1 principle (minL1), the characteristics of which extends beyond ququart systems considered in the figure.

\end{document}